 \documentclass[12pt,preprint,epsfig]{aastex}

 \usepackage[]{graphicx}
 \usepackage{emulateapj5}

 \slugcomment{Accepted version, June 2003}

 \shorttitle{The cosmic evolution of the galaxy luminosity density}
 \shortauthors{Calura \& Matteucci}

\begin{document}


 \title{The cosmic evolution of the galaxy luminosity density}


 \author{Francesco Calura and Francesca Matteucci}
\affil{Dipartimento di Astronomia-Universit\'a di Trieste, Via G. B. Tiepolo
  11, 34131 Trieste, Italy}
 \email{fcalura@ts.astro.it, matteucci@ts.astro.it}




 \begin{abstract}
 We reconstruct the history of the cosmic star formation
 in the universe by means of
 detailed chemical evolution models for galaxies of different morphological 
 types. 
 We consider a picture of coeval, non-interacting evolving
 galaxies where ellipticals experience intense
 and rapid starbursts within the first Gyr after their formation, and spirals 
 and irregulars continue to form stars at lower rates up to the present time.
 Such models allow one to follow in detail the evolution of the metallicity of the gas out of which the
 stars are formed.
 We normalize the galaxy population to the B band luminosity function observed in the local Universe 
 and study the redshift evolution of the luminosity densities in the B, U, I and K bands calculating 
 galaxy colors and evolutionary corrections by means of a detailed synthetic stellar population model. 
 Our predictions indicate that the decline of the galaxy luminosity density 
 between redshift 1 and 0 observed
 in the U, B and I bands is caused mainly by star-forming spiral galaxies which 
 slowly exhaust their gas reservoirs.
 Elliptical galaxies have dominated the total luminosity density in all optical bands 
 at early epochs, when all their stars formed by means of rapid and very intense star-bursts. 
 Irregular galaxies bring a negligible contribution to the total luminosity density in any band 
 at any time. We study the cosmic missing metal crisis and conclude that it could be even more serious
 than what has been assessed by previous authors, if the bulk of the metals were produced in dust-obscured
 starbursts associated 
 with the early spheroid formation.
 The most plausible site for the missing metals could be the warm gas in galaxy groups 
 and proto-clusters, in which the metals could have been ejected through galactic winds following intense
 starbursts. 
 Finally, we predict the evolution of the cosmic star formation and supernova II and Ia rates and obtain
 the best fit to the 
 observations assuming a Salpeter IMF.
 All our results indicate that the bulk of the stellar mass in most galaxies was already in place at early
 epochs and 
 we predict the existence of a peak in the global SFR at high redshift due to ellipticals.

 \end{abstract}




 \section{Introduction}
 The study of the evolution of the galaxy luminosity density is fundamental to 
 understand how galactic structures formed and evolved in the universe.
 The light emitted by stars of various masses at various wavelengths 
 can provide different indications concerning how the cosmic star formation
 history proceeded in the past and concerning the fraction of baryons locked up in
 stars and gas in the local universe.
 The short wavelength light, i.e. that emitted in the rest-frame U and B bands, is mainly radiated
 by short-lived massive stars, thus it can be a direct tracer of star formation.
 On the other hand,
 the near-infrared light is primarily emitted by low mass, long living stars
 which contribute to the bulk of the total stellar mass; hence it traces 
 the mass ditribution of galaxies.
 Several attempts to model the cosmic history of star formation and the evolution of the 
 luminosity density have been performed, generally following different routes.
 One way to reconstruct the evolution of galaxies is often referred to as the ``traditional'' scheme,
 where galaxy densities are normalized according to the optical and IR luminosity functions observed at
 $z=0$.
 The adoption of a star formation history and a cosmological model allows one to follow the redshift
 evolution 
 of the luminosity density
 as well as to reconstruct other properties, such as number counts and the color distributions
 (Totani, Yoshii, Sato 1997; Pozzetti et al. 1998; Tan, Silk \& Balland 1999; Jimenez \& Kashlinsky 1999;
 Totani \& Takeuchi 2002).
 A different approach is followed by
 Madau, Pozzetti \& Dickinson (1998), who start from the observed time-dependent star formation rate (SFR)
 per unit
 comoving volume and an initial mass function (IMF), and then model the luminosity density by means of a
 photometric code.
 Fall, Charlot \& Pei (1996) compute the evolution of the total cosmic emissivity and of the background
 intensity
 on the basis of quasar absorption-line studies, i.e. HI and metallicity data observed in Damped
 Lyman-alpha (DLA)
 systems, combined with cosmic chemical evolution (Pei \& Fall 1995) and stellar population models (see
 also Sadat, Guiderdoni \& Silk 2001
 and Boselli et al. 2001 for similar approaches).
 Other groups compute the cosmic star formation history using large-scale hydrodynamical simulations
 (Nagamine, Cen, 
 Ostriker 2001, Ascasibar et al. 2002) or semi-analytical models of galaxy formation (Baugh et al. 1998,
 Cole et al. 2000, Somerville, Primack, Faber 2001), both methods based on the cold dark matter paradigm.
 An interesting alternative is the approach by Rowan-Robinson (2001), where the star formation rate is
 parametrized in order to 
 investigate the effects that the choice of these parameters has on the far-infrared and submm counts and
 background radiation.\\
 However, most of the approaches described above cannot predict the roles of the different galactic
 morphological types at various cosmic epochs, and which type of galaxy determines the strong evolution 
 observed in the luminosity density between $z=0$ and $z=1$.\\
 In this paper, we calculate the evolution of the galaxy luminosity density and cosmic star formation rate
 by means of detailed 
 chemical evolution models for galaxies of different morphological types, i. e. ellipticals, spirals and
 irregulars, which successfully reproduce the local properties of such galaxies.
 We match our chemical evolution models with a photometric code which allows us to compute the galaxy
 spectra and magnitudes, when the IMF is fixed. 
 We normalize the galaxy fractions according to the B-band luminosity function (LF) as 
 measured by Marzke et al. (1998), and we compute the LF in other bands as a function of redshift on the
 basis of 
 the photometric evolutive corrections and colors predicted by our models.
 Our detailed study allows us to provide an answer to several questions concerning the cosmological
 evolution
 of the baryons in the universe in the form of stars and metals. 
 In particular, we study the cosmic metal production and the 
 related missing metal problem. This can 
 indicate how the chemical enrichment of the inter-galactic medium (IGM)
 has occurred and what might be the nature of the objects observed in the 
 high-redshift universe, such as Lyman-break (LB) and hyperluminous infrared
 galaxies. Finally, we focus on the cosmic evolution of type Ia and II supernova
 (SN) rates, and gain further independent probes of how galactic structures 
 have evolved in the universe since their formation.\\
 The paper is organized as follows: in section 2 we review the current 
 observational status concerning the luminosity density and the star formation 
 measurements in low- and high-redshift galaxies, in section 3 we 
 describe the theory at the basis of our chemo-spectrophotometric models, in 
 section 4 we present our results and in section 5 we draw the conclusions.

 \section{Observations of high-redshift galaxies: integrated luminosity density and cosmic star formation
 rate}
 \subsection{Determining the galaxy luminosity function}
 The total luminosity density (hereinafter LD) in a given band is the integrated light radiated per unit
 volume from the entire galaxy population.
 The distribution of absolute magnitudes for galaxies of any specified Hubble type is represented by the
 luminosity function
 (LF, Efstathiou et al. 1988, Binggeli et al. 1988). 
 The luminosity function is often parametrized according to the form defined
 by Schechter (1976):\\
 \begin{equation}
 \Phi(L)\,dL/L^{\ast} =
 \Phi^{\ast}\,(L/L^{\ast})^{-\alpha}\,exp(-L/L^{\ast})\,dL/L^{\ast}
 \end{equation}\\
 where $\Phi^{\ast}$ is a normalization constant related to the number of luminous galaxies per unit
 volume, 
 $L^{\ast}$ is a characteristic luminosity and $\alpha$ is associated to the slope of the function, which
 accounts for the percentage of faint systems.
 The LD stems from the integral over all magnitudes of the observed luminosity function:\\
 \begin{equation}
 \rho_{L}= \int{\Phi(L)\, (L/L^{\ast})\, dL} 
 \end{equation}\\
 Different determinations of the local field LF (Loveday et al. 1992, Marzke et al. 1994b, Ellis et al.
 1996,
 Marzke et al. 1998, Cross et al. 2001)
 show several discrepancies (Wright 2001), in the sense that all the three 
 parameters are not fully constrained by observations (Ellis 1997).
 Recent progress has being made in the determination of the LF as a function of morphology (Marzke et al.
 1994a, 
 Marzke et al. 1998, Brinchmann et al. 1998, Kochanek et al. 2001) and of spectral type (Heyl et al. 1997,
 Madgwick et al. 2002)
 or color (Blanton et al. 2001), which is important in determining the relative contributions of different
 galaxian
 types to the total luminosity density in the local universe.
 The contributions of different morphological types is a function of the band in which the LF is
 estimated, 
 with early-type galaxies contributing a substantial fraction of the total emissivity at long wavelenths,
 namely in the I and K bands where old massive 
 galaxies dominate (Kochanek et al. 2001), whereas late star-forming spirals are the major contributors in
 the 
 B and U bands (Marzke et al. 1998, Fukugita et al. 1998).
 Notwithstanding the uncertainties in the LF parameters, there is an overall concordance among various
 observations
 regarding the redshift evolution of the galaxy LD.
 This overall agreement concerns, in particular, the $0<z<1$ redshift range, where the LD is observed to
 rise sharply, 
 although the precise steepness is still under debate (Cowie et al. 1999, Lilly et al. 2002). 
 On the other hand, the high redshift trend is rather uncertain, since 
 surface-brightness dimming effects could be significant (Lanzetta et al. 2002)
 and since the bulk of the available data
 has been performed in the rest-frame UV, which can be seriously affected by 
 dust obscuration effects. Dust tends to absorb the UV light emitted by young
 stars and to re-radiate it in the IR-submm bands.
 Moreover, the extent to which the presence of dust 
 contaminates the data is rather uncertain, since galaxies with very 
 different bolometric luminosities can have very similar UV luminosities 
 (Adelberger 2001).
 Some authors (Connolly et al. 1997, Madau et al. 1998) claim a broad peak 
 in the UV LD
 located at $z>1$, followed by a fall at higher redshifts.
 Other authors observe a rather constant behaviour at $z>1$ (Sawicki et al. 1997, Pascarelle et al. 1998),
 which extends to an uncertain epoch when the first galactic structures formed.
 In particular, according to the recent Hubble Deep Field data, the constancy of the galaxy LD 
 is observed out to $z\sim6$, as reviewed in 
 Thompson (2002).

 \subsection{The observed star formation rate density}
 The determination of the SFR density is related to the measures of 
 star formation in galaxies, 
 which can be performed by means of various emission processes (Madau 1997, Kennicutt 1998a, Schaerer
 1999).
 All the different measures require the assumption of a universal mass function,
 which allows the calculation of a multiplicative factor connecting the observed luminosity
to the SFR (for a discussion on the conversion factors used in the literature at various wavelengths, see 
Kennicutt 1998a).
 The UV luminosity density is dominated by the light emitted by young and massive stars, thus it
 provides an estimate of the cosmic SFR density.
 Relevant observations of galaxies in the UV are typically performed at 1500 ${\rm \AA}$ (
 Madau et al. 1996, Madau et al. 1998, Steidel et al. 1999, Massarotti et al. 2001), 
 2000 ${\rm \AA}$ (Treyer et al. 1998) and 2800 ${\rm \AA}$ (Lilly et al. 1996, 
 Sawicki et al. 1997, Connolly et al. 1997, Cowie et al. 1999).
 Other ways to assess star formation include observations of nebular emission lines such as 
 H$\alpha$ (Gallego et al. 1995, Gronwall 1998, Tresse \& Maddox 1998, Glazebrok et al. 1999) and OII 
 (Hammer et al. 1997).
 In all of these cases
 the main source of uncertainty in such determinations is still represented by
 interstellar extinction, caused either by the interstellar medium 
 (ISM) of the observed galaxy or by
 our Galaxy. Star formation activity is likely to take place in highly obscured
 regions, so reliable extinction corrections are required to have accurate estimates of
 the SFR density based on observations in the UV band.\\
 However, without applying any extinction correction, 
 the possibility that very intense star forming regions are heavily 
 obscured by dust 
 implies that observations at short wavelengths can only provide lower limits to the actual SFR density.
 For this reason, the re-processed mid- and FIR- emission from the dust grains,
 which are heated by the UV and optical light emitted by the young stars, is a
 more reliable indicator of star formation activity, and it can be used to
 estimate the dust amount and SFR with fewer selection effects (Blain et al. 1999, Chary \& Elbaz 2001).
 The FIR studies can however be complicated by the multi-component effects of the dust 
 (warm dust, cirrus, Schaerer 1999)
 and by the old stars and AGN contributions to the heating of the dust.
 Observations in the FIR and submm bands include the results by 
 Rowan-Robinson et al. (1997), Hughes et al. (1998) and Flores et al. (1999).
 More recently, observations in the radio continuum have provided other estimates of the
 SFR at high redshift (Mobasher et al. 1999, Haarsma et al. 2000).
 Obviously, the discrepancies seen in the observed evolution of 
 the LD are reflected in the measures 
 of the SFR density, with a globally accepted increasing trend between $z=0$ 
 and $z\sim1$, and a flat behaviour at higher redshifts, 
 though the existence of a peak somewhere at $z>1$ still cannot be completely 
 excluded.

 \section{Chemical and photometric evolution of galaxies}
 By means of chemo-spectrophotometric models of galaxy evolution we aim at reconstructing the history of
 the luminous
 matter in the universe. 
 These models allow us to follow in detail the evolution of the abundances of several chemical
 species, starting from the matter reprocessed 
 by the stars and restored into the ISM through stellar winds and supernova
 explosions. 
We differentiate the galaxy types into ellipticals, spiral disks and irregulars. 
We assume that the category of galactic bulges is naturally included in the one of elliptical galaxies. 
Our assumption is motivated by the fact that they have very similar features: for instance, 
both are dominated by old stellar populations and respect the same fundamental plane (Binney \&
Merrifield 1998). 
This certainly indicates that they are likely to have a common origin, i.e. both are likely to have
formed on very short 
timescales and a long time ago.\\
Detailed descriptions of the chemical evolution models can be found in Matteucci
\& Tornamb\'{e} (1987) and Matteucci (1994) for elliptical galaxies, 
Chiappini et al. (1997) and Chiappini et al. (2001) for spirals and 
Bradamante et al. (1998) for irregular galaxies.
It is worth noting that we choose these models with their specific assumptions 
because they reproduce at best the observed local properties of the various galaxy types.\\
According to our scheme, elliptical galaxies form as the result of the rapid collapse of a homogeneous
 sphere of
 primordial gas where star formation is taking place at the same time as the collapse proceeds, and evolve
 as
 "closed-boxes", i.e. without any interaction with the external environment. 
 Star formation is assumed to halt as the energy of the ISM, heated by stellar winds and SN explosions,
 balances the binding energy of the gas. At this time a galactic wind occurs, sweeping away almost all 
 the residual gas.\\
 Spiral galaxies are assumed to form as a result of two main infall episodes. 
 During the first episode the halo forms and the gas shed by the halo rapidly
 gathers in the center yielding the formation of the bulge. During the second episode, a slower infall
 of external gas forms the disk with the gas accumulating faster in the inner than in the outer
 region ("inside-out" scenario, Matteucci \& Francois 1989). The process of disk formation is much longer
 than the halo
 and bulge formation, with time scales varying from $\sim2Gyr$ in the inner disk to $\sim8Gyr$ in the
 solar region and up 
 to $15-20
 Gyr$ in the outer disk.\\ 
 Finally, irregular dwarf galaxies are assumed to assemble from merging of protogalactic small clouds 
 of primordial composition, until a mass of
 $\sim 6 \times 10^{9}M_{\odot}$ is accumulated, and to produce stars at a lower rate than spirals.
 In the next section, we will present a schematic outline
 of the equations and physical hypotheses at the basis of the models.

 \subsection{The chemical evolution models}
 Let $G_{i}$ be the fractional mass of the element $i$ in the gas
 within a galaxy, its temporal evolution is described by the basic equation:\\
 \begin{equation}
 \dot{G_{i}}=-\psi(t)X_{i}(t) + R_{i}(t) + (\dot{G_{i}})_{inf} - 
 (\dot{G_{i}})_{out}\\
 \end{equation} 
 where $G_{i}(t)=M_{g}(t)X_{i}(t)/M_{tot}$ is the gas mass in the form of an
 element $i$ normalized to a total initial mass $M_{tot}$. The quantity $X_{i}(t)=
 G_{i}(t)/G(t)$ represents the abundance in mass of an element $i$, with
 the summation over all elements in the gas mixture being equal to unity.
 The quantity $G(t)= M_{g}(t)/M_{tot}$ is the total fractional mass of gas
 present in the galaxy at time t.
 $\psi(t)$ is the instantaneous SFR, namely the fractional amount
 of gas turning into stars per unit time; $R_{i}(t)$ represents the returned
 fraction of matter in the form of an element $i$ that the stars eject into the ISM through stellar winds
 and 
 SN explosions; this term contains all the prescriptions regarding the stellar yields and
 the SN progenitor models.
 The two terms 
 $(\dot{G_{i}})_{inf}$ and $(\dot{G_{i}})_{out}$ account for the infalling
 external gas from the intergalactic medium and for the outflow, occurring
 by means of SN driven galactic winds, respectively. 
 The main feature characterizing a particular morphological galactic type is
 represented by the prescription adopted for the star formation history,
 summarized in the SFR expression. 

 In the case of elliptical 
 galaxies the SFR $\psi(t)$ (in $Gyr^{-1}$) has a simple form and is given by:

 \begin{equation}
 \psi(t) = \nu G(t) 
 \end{equation}

 The quantity $\nu$ is the efficiency of star formation, namely the inverse of
 the typical time scale for star formation. In the case of ellipticals, $\nu$ is
 assumed to drop to zero at the onset of a galactic wind, which develops as the
 thermal energy of the gas heated by supernova explosions exceeds the binding
 energy of the gas (Arimoto and Yoshii 1987, Matteucci and Tornamb\'{e} 1987). This quantity is strongly
 influenced by assumptions concerning the presence and distribution of dark
 matter (Matteucci 1992); 
 for the model adopted here a diffuse 
 ($R_e/R_d$=0.1, where
 $R_e$ is the effective radius of the galaxy and $R_d$ is the radius 
 of the dark matter core) but 
 massive ($M_{dark}/M_{Lum}=10$) dark halo has 
 been assumed.\\
 In the case of irregular galaxies we have assumed a continuous star formation rate always expressed as in
 (4), but
 characterized by an efficiency lower than the one adopted for ellipticals.\\
 In the case of spiral galaxies, the SFR expression (Chiappini et al. 1997) is:\\
 \begin{equation}
 \psi(r,t) = \nu [\frac{\sigma(r,t)}{\sigma(r_{\odot},t)}]^{2(k-1)}
 [\frac{\sigma(r,t_{Gal})}{\sigma(r,t)}]^{k-1}
 \sigma_{gas}(r,t)
 \end{equation}
 \\
 where $\nu$ is the SF efficiency, $\sigma(r,t)$ is the total (gas + stars) surface mass density at a
 radius r and time t,
 $\sigma(r_{\odot},t)$ is the total surface mass density in the solar region and $\sigma_{gas}(r,t)$ is
 the surface gas 
 density. For the gas density exponent $k$ a value of 1.5 has been assumed by Chiappini et al. (1997) in
 order to ensure a good
 fit to the observational constraints at the solar vicinity and in agreement with the estimates by
 Kennicutt (1998b).
 The three different star formation rates for ellipticals, spirals and irregulars as functions of time and
 for a Salpeter IMF 
 are shown in figure 1.

 \begin{figure*}
 \centerline{\includegraphics[height=19pc,width=19pc]{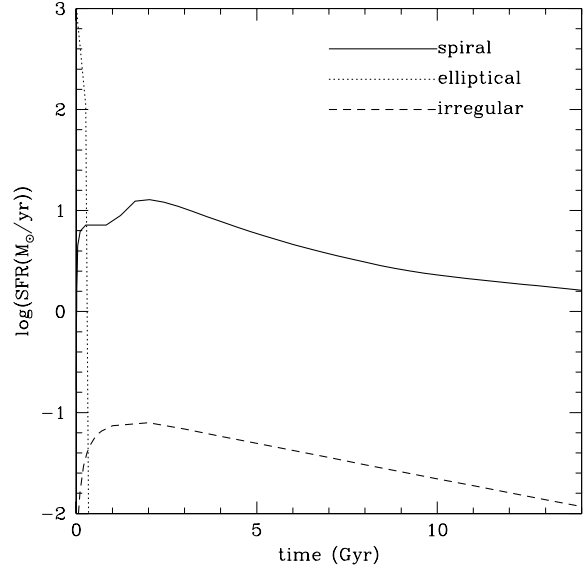}} 
 \caption[]{Star formation rates as functions of time for a spiral, elliptical and irregular galaxy in the
 case of a Salpeter IMF 
 calculated according to equations (4) and (5).
 }
 \end{figure*}

 \subsubsection{The Supernova Rates}
 The true nature of Type Ia SNe and their progenitors is currently a matter of debate.
 Several models have been proposed so far, but divided between two main competing scenarios.
 According to the single degenerate (SD, Whelan \& Iben 1973) scenario, a C-O white dwarf accretes mass
 from a non-degenerate companion until it reaches the Chandrasekhar mass ($\sim 1.4 M_{\odot}$) 
 and explodes via C-deflagration leaving no remnant.
 The alternative double-degenerate (DD, Iben \& Tutukov 1984) model sees the merging of two C-O white
 dwarfs which,
 owing to the loss of angular momentum via gravitational wave radiation, coalesce and explode by
 C-deflagration.
 For our purposes we adopt the SD model, which generally best represents the characteristics of the
 majority of type Ia SNe, 
 as discussed in Matteucci \& Recchi (2001).
 Type II SNe explosions are supposed to originate from core collapse of single massive ($M > 8 M_{\odot}$)
 stars,
 with maximum masses allowed of $100 M_{\odot}$. 

 The type Ia SN rate is expressed as:
 \begin{equation}
 R_{Ia}(t) = A \int_{M_{Bm}}^{M_{BM}} \phi(M_{B}) [\int_{\mu_{m}}^{0.5} f(\mu)
 \psi(t-\tau_{M_{2}})d\mu]dM_{B}
 \end{equation}
 where A is the fraction of binary systems which can end as SNIa within the IMF mass range, and which have
 a total mass
 $M_{Bm}\le M_{B} \le M_{BM}$.
 $\mu=M_{2}/M_{B}$ is the ratio between the secondary component of the binary system (i.e. the less
 massive one) and the total mass of the system and $f(\mu)$ is the 
 distribution function of this ratio. Statistical studies indicate that mass 
 ratios close to $0.5$ are preferred, so the formula:\\
 \begin{equation}
 f(\mu)=2^{1+\gamma}(1+\gamma)\mu^{\gamma}
 \end{equation}\\
 is commonly adopted (Matteucci \& Recchi 2001), with $\gamma=2$ as a parameter.
 $\tau_{M_{2}}$ is the lifetime of the less massive star in the
 system, which lives longer and therefore determines the time-scale for the explosion.
 The assumed value of A is $\sim 0.1$. This value assures that the present time SN rates 
 are reproduced in each galactic type.
 For the masses ${M_{Bm}}$ and ${M_{BM}}$ we
 chose the values $3M_{\odot}$ and $16M_{\odot}$, respectively (see Matteucci \& Greggio 1986).\\ 
 The type II SN rate is simply expressed as:\\
 \\
 \begin{math}
 R_{II}(t) = (1-A) \int_{>8}^{M_{BM}} \phi(M) \psi(t-\tau_{M})dM + 
 \end{math}
 \begin{equation}
 \int_{M_{BM}}^{100} \phi(M) \psi(t-\tau_{M})dM 
 \end{equation}
 \\
 The IMF has been assumed to be constant in space and
 time and among galaxies. We have compared results calculated with two different IMFs, i. e. 
 the Salpeter (1955) and the Scalo (1986).\\

 \subsection{The spectro-photometric model}

 The spectro-photometric calculations have been performed by means of the model by Jimenez et al. (1998).
 This model is based on the stellar isochrones computed by Jimenez et al. (1998) and the stellar
 atmospheric models
 by Kurucz (1992). 
 The main advantage of this photometric code is that it allows one to follow in detail at every single
 timestep
 the metallicity evolution of the gas out of which the stars form, at variance with other popular
 photometric models which
 require the assumption of a constant metallicity.
 Using the stellar inputs, first we build simple stellar population (SSP) models consistent with the
 chemical evolution 
 at any given time and 
 weighted according to the assumed IMF. Then, a composite stellar population (CSP) consists of the sum of
 different SSP formed 
 at different times, with a luminosity at an age $t_{0}$ and at a particular wavelength $\lambda$ 
 given by:\\

 \begin{equation}
 L_{\lambda}(t_{0})=\int_{0}^{t_{0}} \int_{Z_{i}}^{Z_{f}} \psi(t-t_{0}) L_{SSP,\lambda}(Z,t-t_{0})dZdt
 \end{equation}

 where the luminosity of the SSP can be written as:\\

 \begin{equation}
 L_{SSP,\lambda}(Z, t_{0}-t)= \int_{M_{min}}^{M_{max}} \phi(m) l_{\lambda}(Z,M,t_{0}-t)dM
 \end{equation}\\

 and $l_{\lambda}(Z,M,t_{0}-t)$ is the luminosity of a star of mass M, metallicity Z and age $t_{0}-t$;
 $Z_{i}$ and $Z_{f}$ are the initial and final metallicities, $M_{min}$ and $M_{max}$ are the smallest and
 largest 
 stellar mass in the population, $\phi(m)$ is the IMF and $\psi(t)$ is the SFR at the time t. 

 \section{Results}

 \subsection{The evolution of the galaxy luminosity density} \label{bozomath}
 In the B band, at $z=0$ the LDs for the single galaxy types are simply given by the integral of the LFs
 observed by Marzke et al. 
 (1998, see their equation 2). At redshift other than zero we tranform the absolute magnitudes applying
 the evolutionary corrections, 
 calculated by means of the spectro-photometric code for every galaxy type:
 \begin{equation}
 M_{B}(z)=M_{B}(z=0)+2.5log(\frac{\int E_{\lambda/1+z}(z)R_{B}(\lambda)d\lambda}{\int
 E_{\lambda/1+z}(0)R_{B}(\lambda)
 d\lambda})
 \end{equation}\\
 where $M_{B}(z=0)$ and $M_{B}(z)$ are the absolute blue magnitudes at redshift 0 and z, respectively, 
 $E_{\lambda}(z)\, d\lambda$ is the energy per unit time radiated at the rest-frame wavelength $\lambda$
 by 
 the galaxy at redshift $z$, 
 and $R_{B}(\lambda)$ is the response function of the rest-frame B band. 
 The second term on the right side of equation (11) represents the evolutionary correction (EC), i.e. the
 difference in absolute 
 magnitude measured in the rest frame of the galaxy at the wavelength of emission (Poggianti 1997).\\
 We then calculate the B band LF at redshift z according to:\\
 \begin{equation}
 \Phi_{B}(M_{B},z)=\Phi_{B}(M_{B}(z))
 \end{equation}\\ 
which is equivalent to assuming evolution in luminosity and not in number. 
This clearly coincides to assume that the effects of mergers are small at all redshifts.
 In bands other than B we assume that the LF shape is the same as in the B band and we 
 calculate the LF in the given band (X) transforming the absolute magnitudes according to the 
 rest-frame galaxy colors 
 as predicted by the spectrophotometric model:\\
 \begin{equation}
 M_{X}=M_{B}+(X-B)_{rf}
 \end{equation}\\
 Finally, the total luminosity density per unit frequency in a given band $\rho_{\lambda}$ is
 given by the sum of the single contributions of the three different galactic 
 morphological types (ellipticals, spirals and irregulars):\\
 \begin{equation}
 \rho_{\lambda}(z) =\sum_{i} \rho_{\lambda,i}(z)
 \end{equation}\\
 Figures 2 and 3 show the predicted evolution of the total LD in the U (centered at 3650${\rm \AA}$), 
 B (centered at 4450 ${\rm \AA}$), I (centered at 8060 ${\rm \AA}$) and K (centered at 21900 ${\rm \AA}$) 
 bands and the contributions
 of each morphological type for a Salpeter (1955) and a Scalo (1986) IMF, respectively.
 In all bands, at early times the total LD is dominated by the light produced by elliptical galaxies,
 which experience their strong star-burst during which they are the most intense sources in each 
 photometric band. After this phase, which lasts $\sim 0.3$ Gyr, they will evolve passively for
 the rest of their life. On the other hand, 
 the disks of spiral galaxies form stars continuously at all epochs: their luminosity increases slowly until $3Gyr$
 from the 
 beginning of the star formation, 
 where it reaches a broad peak, then it starts to decrease
 significantly, owing to the progressive consumption of their gas reservoirs.\\
 Irregular galaxies are the most slowly evolving systems and their star formation rates never reach the
 values
 recorded in spiral disks and ellipticals. 
 As a consequence, their LD values are the smallest at any time and in any band.

 \begin{figure*}
 \centerline{\includegraphics[height=19pc,width=19pc]{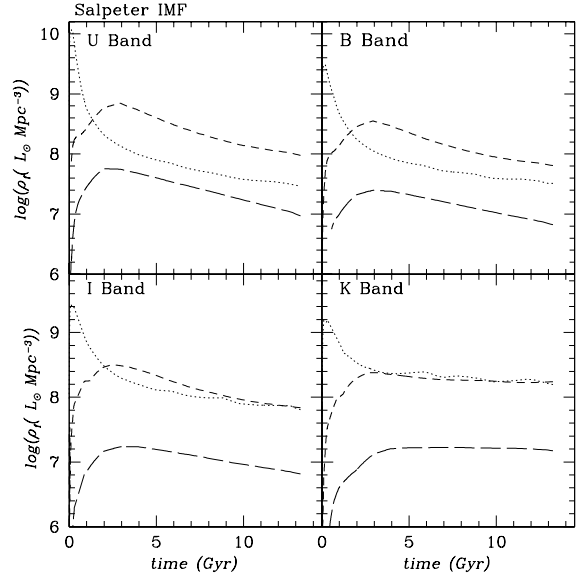}} 
 \caption[]{Predicted evolution of the luminosity density for galaxies of different types in the U, B, I
 and K bands
 for a Salpeter IMF.
 Dotted line: elliptical galaxies; short-dashed line: spiral galaxies; long-dashed line: irregular
 galaxies.
 }
 \end{figure*}

 \begin{figure*}
 \centerline{\includegraphics[height=19pc,width=19pc]{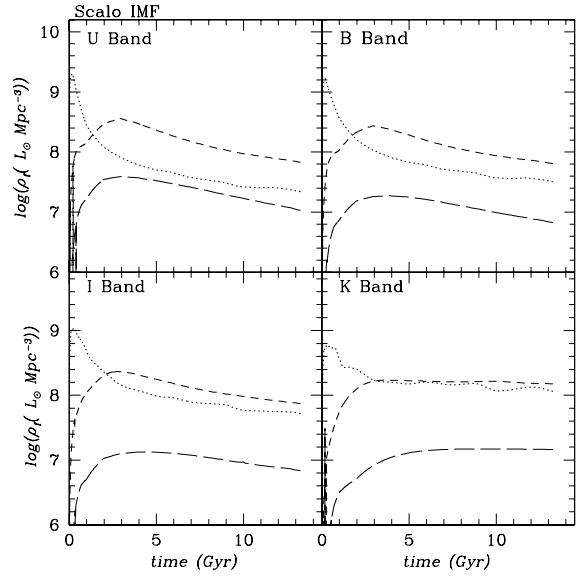}} 
 \caption[]{Predicted evolution of the luminosity density for galaxies of different types in the U, B, I
 and K bands
 for a Scalo IMF.
 Dotted line: elliptical galaxies; short-dashed line: spiral galaxies; long-dashed line: irregular
 galaxies.
 }
 \end{figure*}

 In the I and K bands the difference between the spiral and elliptical luminosity densities are small
 after 3 Gyr.
 At this epoch the total amount of stars is nearly in place in both types.
 At the present time (13 Gyr), the contributions of ellipticals and spirals to the total luminosity
 density in the I and K bands
 is practically the same, at variance with what happens in the B and U bands.
 The difference between the Salpeter and the Scalo IMF concerns the stars in the high mass regime
 ($M>8M_{\odot}$), which 
 are more numerous in the former case. This translates into the amount of UV and B light emitted by all
 galaxies, 
 which is sensitively larger with the Salpeter than with the Scalo IMF.
 At longer wavelengths the difference diminishes: in the I band, where there is still a significant
 contribution brought 
 by high intermediate mass stars experiencing the asymptotic giant branch (AGB) phase, the difference
 between the elliptical and
 spiral luminosity is stronger with the Scalo IMF, whereas the K band luminosities have practically the
 same behaviour
 in both cases.

 Figure 4 shows the predicted and observed evolution of the LD in various bands for a Salpeter (left
 panels)
 and a Scalo (right panels) IMF and for a galaxy formation redshift of $z_{f}=5$, 
 having assumed an Einstein-De Sitter (EdS) cosmological model ($\Omega_{0}=1$, $\Omega_{\Lambda}=0$) and
 $h=0.5$. 
 Figure 5 shows the same as in figure 4 but in the case of a Lambda-cold dark matter cosmology
 ($\Lambda$CDM, $\Omega_{0}=0.3$, 
 $\Omega_{\Lambda}=0.7$) and $h=0.65$. 
 Figures 6 and 7 show the same as in figures 4 and 5, respectively, but with a galaxy formation redshift
 of $z_{f}=10$.
 The data have been converted from the EdS to the $\Lambda$CDM cosmology according to the
 prescriptions given by Somerville et al. (2001).
 Given the slight discrepancy between the B-band LF normalization by Marzke et al. (1998), adopted for our
 estimate in the local
 universe, and the one by Ellis et al. (1996), which is shown in the plots, 
 we have re-normalized our value at $z=0$ to the one by Ellis et al. (1996) in order to better stress
 the goodness of our fit.

 \begin{figure*}
 \centerline{\includegraphics[height=22pc,width=35pc]{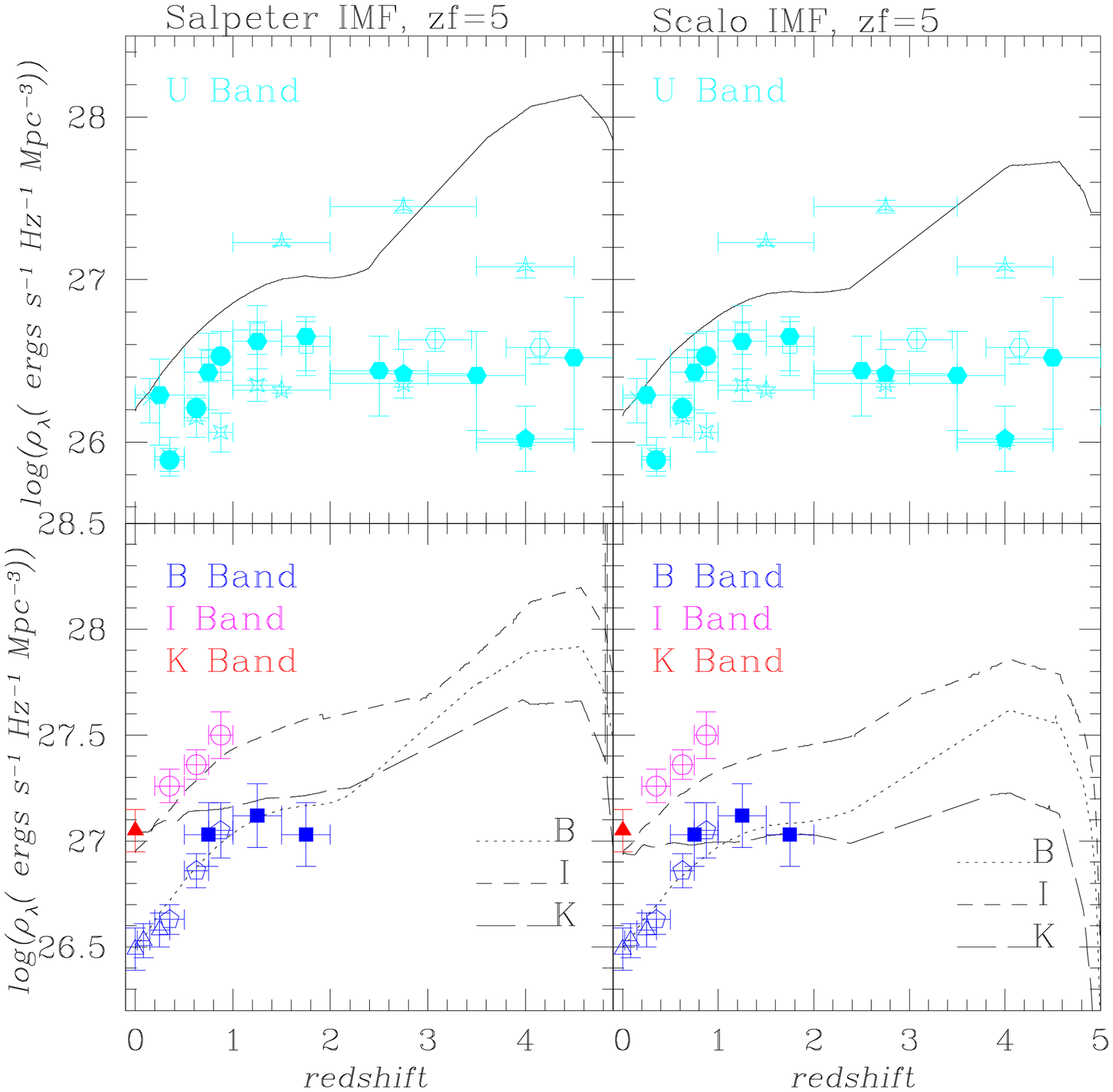}} 
 \caption[]{
 \emph{Upper Panels}: predicted (solid line) and observed luminosity densities in the U band for a
 Salpeter (left panel) and Scalo 
 (right panel) IMF in the 
 case of galaxy formation at $z_{f}=5$.
 \emph{Lower panels}: predicted and observed luminosity densities in the B, I, K bands.
 The adopted cosmology is the one of Einstein-De Sitter ($\Omega_{0}=1$, $\Omega_{\Lambda}=0$) with
 $h=0.5$. 
 Data in the upper panels:
 Solid hexagons from Pascarelle et al. (1999, $1500$ ${\rm \AA}$), 
 open hexagons from Steidel et al. 1999,
 cross from Treyer et al. (1998, $2000$ ${\rm \AA}$),
 stars from Massarotti et al. (2001, $1500$ ${\rm \AA}$), 
 solid pentagons from Madau et al. (1998, $1500$ ${\rm \AA}$), 
 solid circles from Lilly et al. (1996, $2800$ ${\rm \AA}$), 
 open squares from Connolly et al. (1997, $2800$ ${\rm \AA}$,
 four-points stars from Cowie et al. 1999.
 The three-points stars are the extinction-corrected data by Massarotti et al. (2001).

 Data in the lower panels:
 Open triangles from Ellis et al. (1996, $4400$ ${\rm \AA}$, B), 
 open pentagons from Lilly et al. (1996, $4400$ ${\rm \AA}$, B), 
 solid squares from Connolly et al. (1997, $4400$ ${\rm \AA}$, B), 
 open circles from Lilly et al. (1996, $1$ $\mu$, I),
 solid triangle from Gardner et al. (1997, $2.2$ $\mu$, K). 
 }
 \end{figure*}

 \begin{figure*}
 \centerline{\includegraphics[height=22pc,width=35pc]{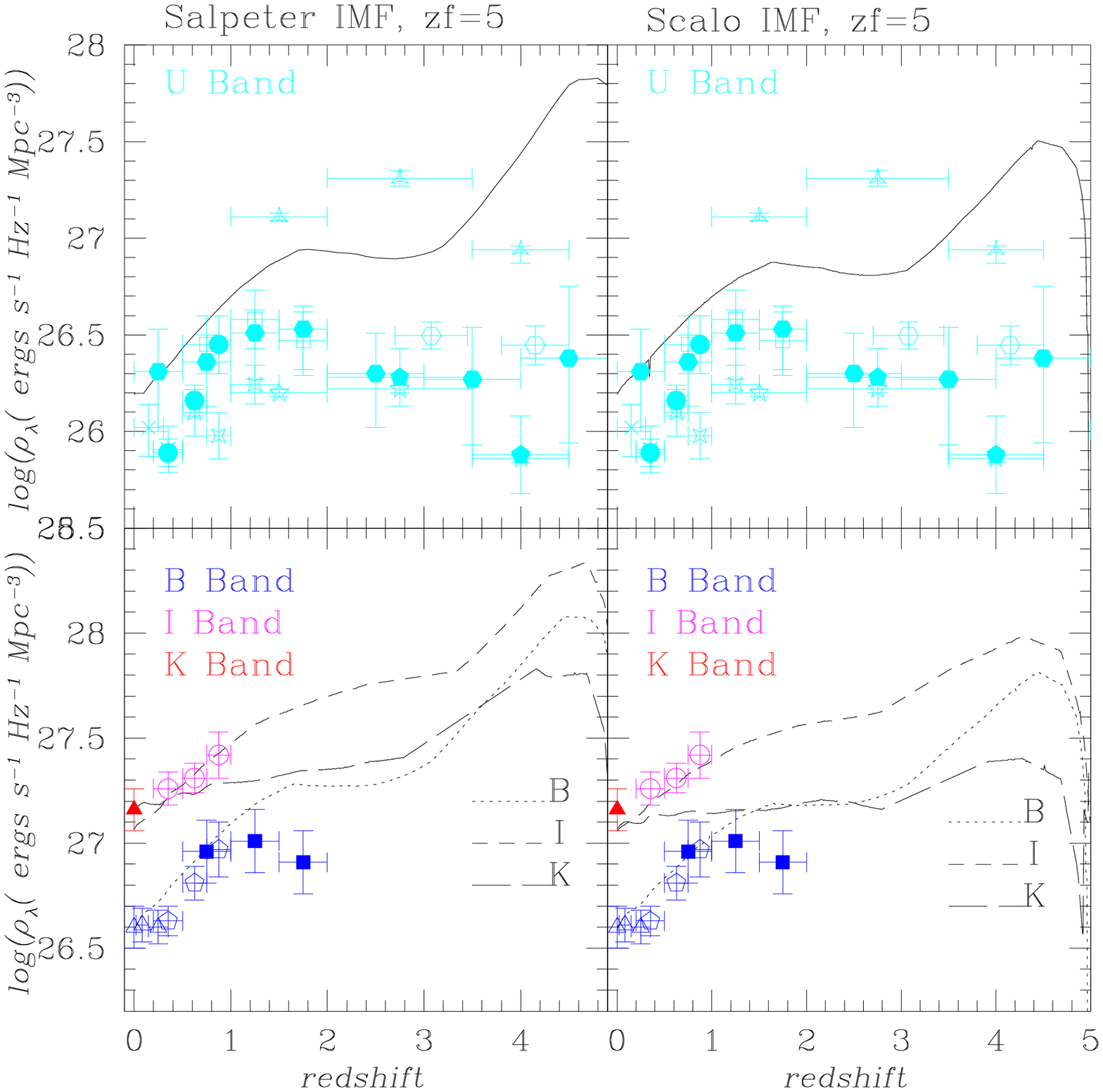}} 
 \caption[]{
 \emph{Upper Panels}: predicted (solid line) and observed luminosity densities in the U band for a
 Salpeter (left panel) and Scalo 
 (right panel) IMF in the 
 case of galaxy formation at $z_{f}=5$.
 \emph{Lower panels}: predicted and observed luminosity densities in the B, I, K bands.
 The adopted cosmology is the $\Lambda$CDM ($\Omega_{0}=0.3$, $\Omega_{\Lambda}=0.7$) with $h=0.65$.
 Data are as described in figure 4.
 }
 \end{figure*}

 \begin{figure*}
 \centerline{\includegraphics[height=22pc,width=35pc]{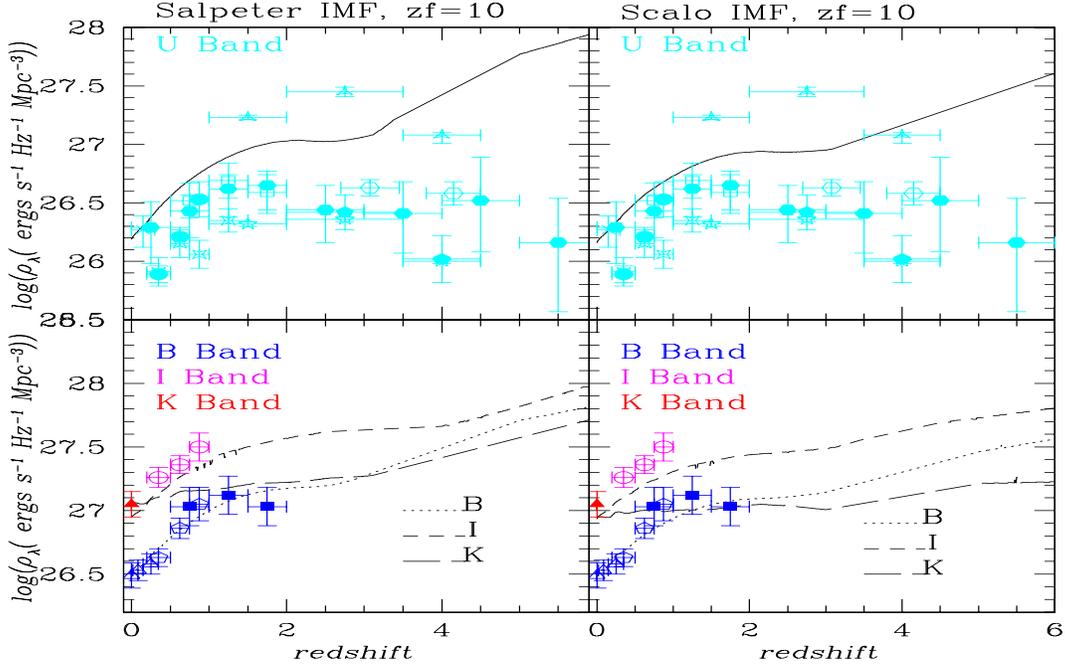}} 
 \caption[]{ 
 \emph{Upper Panels}: predicted (solid line) and observed luminosity densities in the U band for a
 Salpeter (left panel) and Scalo 
 (right panel) IMF in the 
 case of galaxy formation at $z_{f}=10$.
 \emph{Lower panels}: predicted and observed luminosity densities in the B, I, K bands.
 The adopted cosmology is the one of Einstein-De Sitter ($\Omega_{0}=1$, $\Omega_{\Lambda}=0$) with
 $h=0.5$. 
 Data are as described in figure 4.
 }
 \end{figure*}

 \begin{figure*}
 \centerline{\includegraphics[height=22pc,width=35pc]{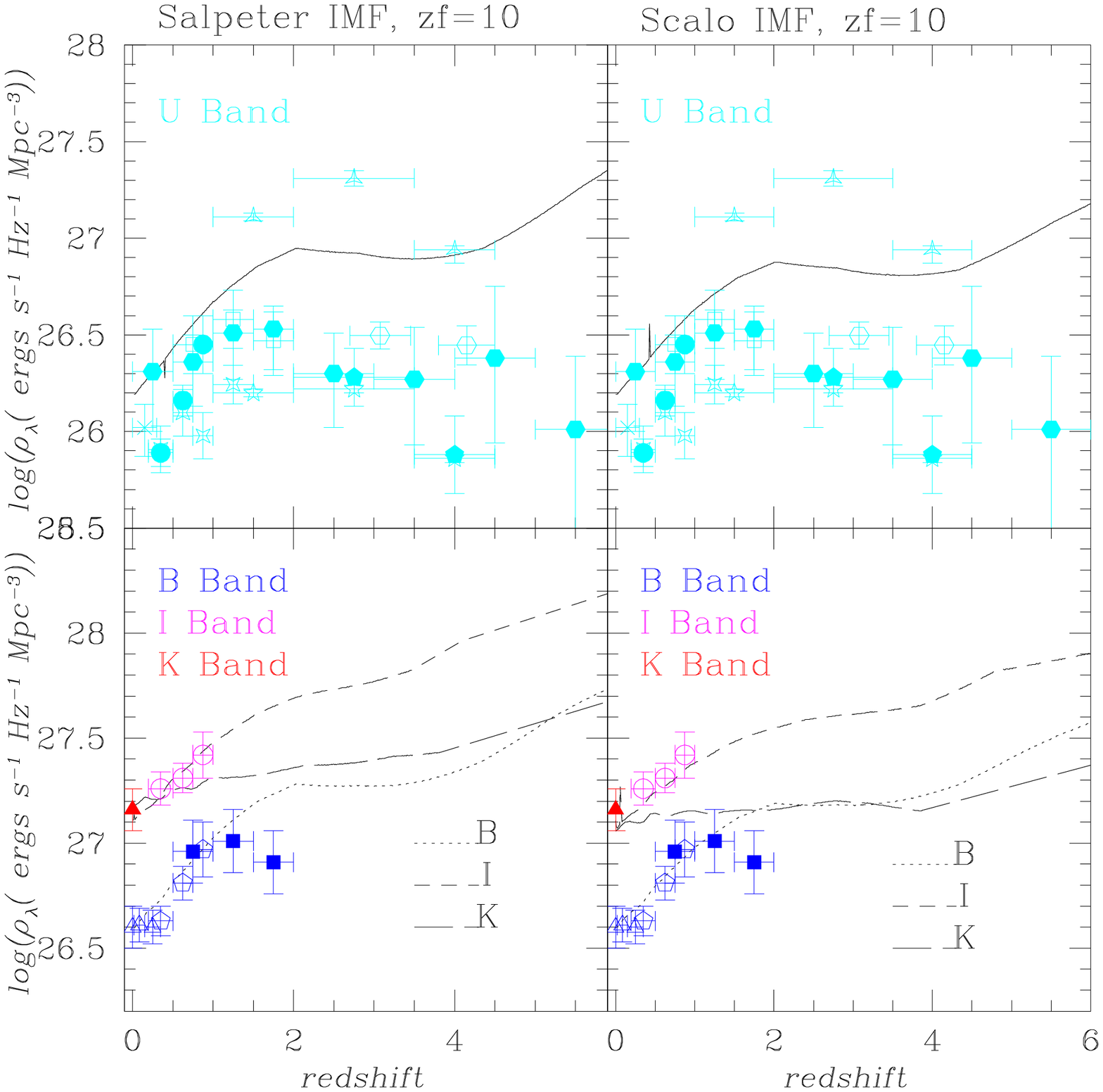}} 
 \caption[]{
 \emph{Upper Panels}: predicted (solid line) and observed luminosity densities in the U band for a
 Salpeter (left panel) and Scalo 
 (right panel) IMF in the 
 case of galaxy formation at $z_{f}=10$.
 \emph{Lower panels}: predicted and observed luminosity densities in the B, I, K bands.
 The adopted cosmology is the $\Lambda$CDM ($\Omega_{0}=0.3$, $\Omega_{\Lambda}=0.7$) with $h=0.65$.
 Data are as described in figure 4.
 }
 \end{figure*}

 In the U band (upper panels) the data show a considerable spread, with large error bars at any redshift.
 All the UV data reported in the figure are not extinction-corrected, with the only exception of the three
 points
 measured by Massarotti et al. (2001).
 Our predictions indicate a peak in the total UV LD corresponding to the redshift of galaxy formation and
 a 
 monotonically decreasing trend down to $z=0$, with a particularly strong evolution between $z=1$ and
 $z=0$.
 This last feature is 
 in agreement with the data by Pascarelle et al. (1998) and with the low redshift value by Treyer et al.
 (1998).
 We do not see evidence for the very steep drop observed by Lilly et al. (1996).
 From figures 4 and 5 we note that, if the peak generated by elliptical formation were shifted at
 $z\sim3$, it would be fairly
 consistent with the extinction-corrected measures by Massarotti et al. (2001), in particular in the case
 of the Scalo
 IMF. To our knowledge, the extinction-corrected data by Massarotti et al. (2001) represent the first
 observational 
 evidence for a very high peak in the observed UV LD which could be clearly ascribed to the massive star
 formation 
 occurring during the formation of spheroids.\\
 In the B-band (lower panels) there is a smaller number of detections than in the UV but a better
 agreement between the observations
 and the predictions in the $0<z<1$ redshift range.
 The smallest number of observations is found in the I and K bands. At $z=0$ the K-band LD is rather well
 constrained by several 
 measures (Gardner et al. 1997, Loveday 2000, Cole et al. 2001, Kochanek et al. 2001), which show a
 substantial agreement with one another.
 For clarity, in our plots we have chosen one single value, the one by Gardner et al. (1997).
 At redshift larger than zero, in the I band the only observations at our disposal are the ones by Lilly
 et al. (1996) at $1$ $\mu$. 
 We note that our curves reproduce very well the oserved evolution of the LD in the B, I and K bands.
 The best agreement between the data and the predictions is achieved in the case of the $\Lambda$CDM
 cosmology,
 especially in the I and K bands. This occurs
 mainly because, having 
 assumed galaxy formation redshifts of $z_{f}=5$ and $z_{f}=10$, 
 in the $\Lambda$CDM cosmology a formation redshift of $z_{f}=5$ ($z_{f}=10$) corresponds to a lookback
 time of 13.26 Gyr
 (14 Gyr), whereas in the EDS $z_{f}=5$ ($z_{f}=10$) corresponds to 12.15 Gyr (12.68 Gyr).
 This means that in the $\Lambda$CDM galaxies are older than in the EDS, consequently they contain
 redder stellar populations. 
 Both the Salpeter and the Scalo IMFs provide satisfactory results.
 As already mentioned, the Salpeter IMF is richer in high mass stars than the Scalo IMF.
 The consequence of this is that 
 with the Salpeter IMF all galaxies are more luminous at the shortest wavelengths, 
 where these stars emit the bulk of their light.
 Moreover, with the Salpeter IMF galaxies are on average redder than with the Scalo one.
 This translates into values of the I and K LDs slightly higher. However, given the
 uncertainties and the small data sample in the I and K bands, it is difficult to conclude 
 which choice for the IMF should be preferred.
 \emph{Our main result is that the disks of spiral galaxies are mainly responsible for the observed decrease of the
 total LD in the U 
 and B bands between $z\sim1$ and $z\sim0$, whereas in the I band the decline is due to contributions from
 both 
 spirals and ellipticals, as can be seen also in figures 2 and 3.}

 Figures 6 and 7 show the evolution of the comoving LD in various bands 
 assuming that galaxies formed at $z_{f}=10$, having assumed an EdS and 
 $\Lambda$CDM cosmology, respectively, and in the cases of a Salpeter
 (left panels) and a Scalo (right panels) IMF.
 The predictions are still in fair agreement with the observations, in particular
 in the $0<z<1$ redshift range. 
 The main differences concern the predictions at 
 high redshift: the $z_{f}=10$ scenarios allow for a nearly constant LD throughout a wider redshift range.
 As far as the $z<1$ behaviour of the LD is
 concerned, little difference can be seen when we assume a global galaxy formation
 at $z_{f}=5$ or $z_{f}=10$. 
 This is due to the short cosmic time 
 elapsed between $z=5$ and $z=10$ both in the EdS and $\Lambda$CDM cosmologies,
 namely $0.53$ and $0.74$ Gyr, respectively. This time scale is much shorter than 
 the one corresponding to the $0<z<1$ redshift range, i. e. $8.43$ Gyr in the EdS and $8.31$ Gyr
 in the $\Lambda$CDM, having thus little influence on what happens during the
 latter period.\\
 The study of the galaxy LD at high redshift is crucial for understanding the 
 epoch of spheroid formation. Unfortunately, the high redshift measures concern only
 the UV band, which is the most seriously affected by dust extinction effects
 (Adelberger \& Steidel 2000).
 If the attenuation by dust is as serious as advocated by Massarotti et al. 
 (2001), the peak generated by the formation of massive spheroids could lie 
 anywhere in the $z>1$ region, but it would be hidden if the
 associated strong starbursts occurred in sites heavily obscured by dust.
 In fact, spheroids are the systems which most rapidly reach oversolar metallicities 
 (Pettini et al. 2002, Matteucci \& Pipino 2002), therefore the formation of dust grains could be
 particularly favoured if,
 as is reasonable to assume, the probability of dust formation is proportional to the metal content.
 High redshift observations at higher wavelengths, i. e. in bands not 
 affected by biases caused by dust extinction,
 could provide fundamental hints on the epoch of major spheroid formation, 
 as well as high redshift observations in the FIR/sub-mm bands, where dust
 re-emits all the starlight absorbed in the rest-frame B and U bands 
 (Hughes et al. 1998, Blain et al. 1999).

 \subsection{The cosmic star formation rate density} \label{bozomath}
 Figure 8 shows the evolution of the cosmic SFR density as a function of redshift (Madau's plot) 
 as predicted by our models and as observed 
 by several authors in the case of a $\Lambda$CDM cosmology and galaxy formation at $z_{f}=5$.
 The cosmic SFR density is not a directly observed quantity: to be evaluated it needs, besides 
 the measure of the LD at certain wavelengths (mostly $1500$ and $2000 {\rm \AA}$), 
 the adoption of a universal IMF
 to calculate the conversion factor between the observed $\rho_{\lambda}$ and $\dot{\rho_{*}}$.
 For this reason the determination of this quantity is affected by two main sources of uncertainty.
 One is related to the determination of the calibration constant, uncertain if the universal IMF is
 varying 
 throughout cosmic time. The second is related to the dust extinction affecting the UV observations.
 We have computed the cosmic SFR density according to:\\
 \begin{equation}
 \dot{\rho_{*}}(z) = \sum_{i} \rho_{B i}(z) \, (\frac{M}{L})_{B i}(z) \, \psi_{i}(z)
 \end{equation} \\
 where $ \rho_{B i}$, $(\frac{M}{L})_{B i}$ and $\psi_{i}$ are the B LD, the B mass-to-light ratio and
 star formation 
 rate for the galaxies of the $i$-th morphological type, respectively.
 The points in figure 8 represent estimates based on observations at short wavelengths, except the solid
 hexagon 
 obtained by means of sub-mm measures by Hughes et al. (1998). 
 However, since the extent of the attenuation by dust is highly uncertain (see Steidel 
 et al. 1999, Hopkins et al. 2001, Thompson et al. 2001), we have chosen to use the data uncorrected for
 dust extinction. For this reason, all the data based on optical observations should be regarded as lower
 limits 
 to the true values.
 The thick lines in the curves indicate lower and upper limits on the SFR density
 calculated by Haarsma et al. (2000) from radio observations. 
 As well as the point from sub-mm data, these estimates need no corrections for 
 dust obscuration since the radio emission at $\nu > 1 GHz$ passes freely through dust.\\ 
 At $z\sim0$ we overpredict the value observed by Gallego et al. (1995) by a factor of 4. 
 Interestingly, our estimate is instead in very good agreement with a recent determination based on Local
 Group observations by 
 Hopkins, Irwin and Connolly (2001), which calculate a local SFR density higher than the one by Gallego et
 al. (1995) 
 by a factor of $\sim 5$. 
 This fact implies that, even in the local universe, dust obscuration could introduce serious biases in
 the 
 determinations of the global SFR density. 
 Since the probability to intercept dusty objects increases in lockstep with the line of sight, 
 it is conceivable that this effect could become more and more pronounced at increasing redshift.
 Another source of discrepancy between our local value and the one by Gallego et al. (1995) is 
 the normalization of the luminosity density that we have chosen, which in this case seems too high. 
 Since the discrepancy between our local value and the observed one arises from the combination of both
 effects, 
 it seems very difficult to quantify the uncertainty associated to the choice of the normalization. 
 If we were to neglect all the effects due to dust obscuration, and if we took the value by Gallego et al.
 (1995) at face value, 
 the uncertainty due to the normalization would be of a factor of 4.
 In figure 8, in order to better stress the comparison between our prediction 
 and the observations, our curve has been re-normalized to the 
 value by Gallego et al. (1995). 
 Our predictions indicate a peak in the SFR density at high redshift due to massive star formation in
 spheroids. 
In our models, the SFR in ellipticals can reach very high values (up to $1000 M_{\sun}/yr$), very similar 
to the ones observed in some SCUBA (Ivison et al. 2000) and Hyperlouminous infrared galaxies 
 (Rowan-Robinson et al. 1997, Rowan-Robinson 2000). 
 Such values are motivated by the fact that, in order to reproduce several observational features
 such as the observed increase of the [Mg/Fe] ratio and the velocity dispersion $\sigma$, 
 giant elliptical galaxies, 
 i.e. with masses of the order of $\sim 10^{11} M_{\odot}$, have to form all their stars on very short
 timescales, 
 i.e. $\tau \le 1$ Gyr (Matteucci 1994). Hence, the possible SFR rates for high mass ellipticals 
 span the range $\sim 100 - 1000 M_{\odot}/yr$.
 If we were to assume a galaxy of $M \sim 10^{10} M_{\odot}$ as a typical elliptical, with a star
 formation timescale of $\sim 1$ Gyr, 
 this would lower the peak in the SFR density by a factor of $10$.
 As already mentioned in section 4.1, if most of the star formation in the universe occurred at very high
 redshift 
 and in sites highly obscured by dust, there could be a peak somewhere at $z>1$ which would remain
 completely unseen in the observed $\dot{\rho_{*}}$-z plot.

 \begin{figure*}
 \centerline{\includegraphics[height=22pc,width=30pc]{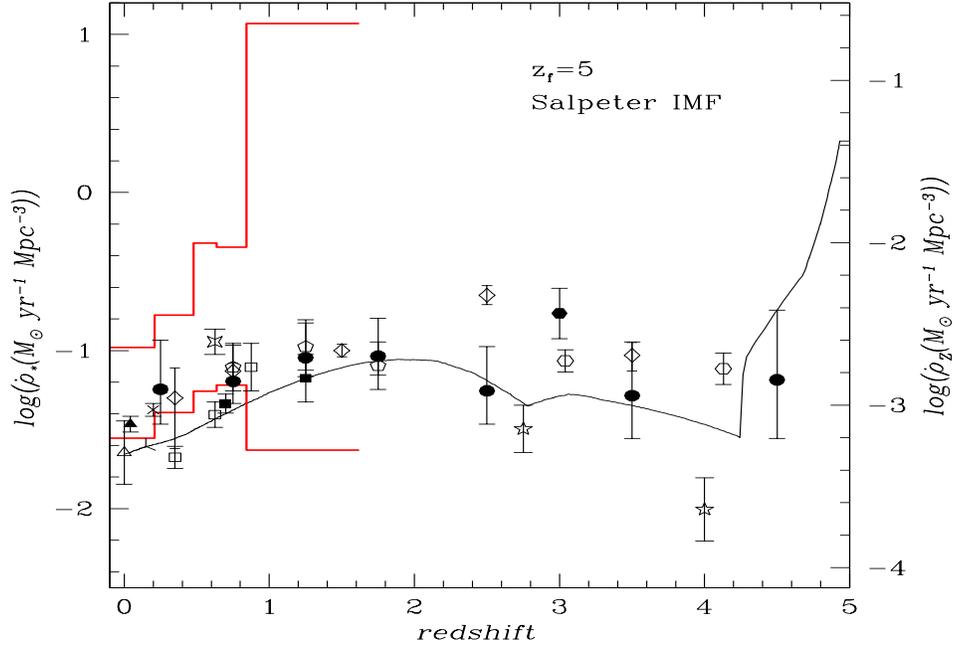}} 
 \caption[]{Predicted and observed star formation rate density (left scale) and metal-ejection rate
 density (right scale)
 for a Salpeter IMF in the case of a $\Lambda$CDM cosmology and galaxy formation at $z_{f}=5$.
 All the points are taken from Somerville et al. 2001. The thick lines represent lower and upper limits
 based on radio 
 observations (Haarsma et al. 2000).}
 \end{figure*}

 \begin{figure*}
 \centerline{\includegraphics[height=22pc,width=30pc]{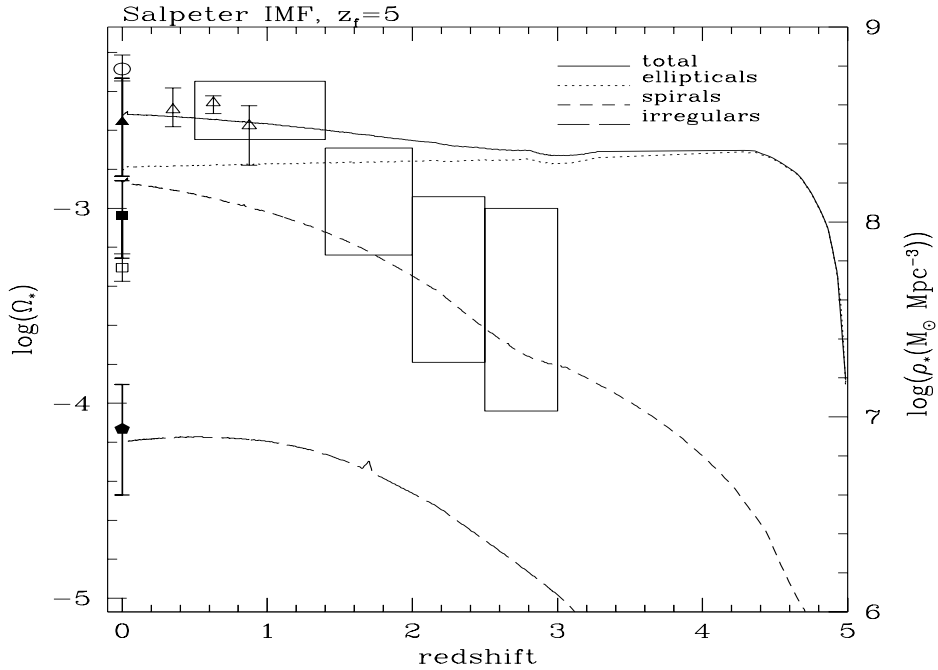}} 
 \caption[]{Predicted and observed comoving stellar mass density $\Omega_{*}$ 
 for a $\Lambda$CDM cosmology, a Salpeter IMF and galaxy formation occurred at $z_{f}=5$.
 Dotted line: predicted evolution of the stellar mass density 
(living stars plus remnants) in elliptical 
 galaxies; short-dashed line: predicted contribution from spiral galaxies; 
 long-dashed line: predicted contribution from irregular galaxies; 
 solid line: predicted total evolution of the stellar mass density. 
The boxes represent the values observed by Dickinson et al. (2003), 
the open circle is the $z=0$ value observed by Cole et al. (2001),   
the open triangles are by Brinchmann \& Ellis (2000) 
and the open square is the value observed by P\'erez-Gonz\'alez et al. (2003) in local star-forming galaxies. 
The solid triangle, solid square and solid pentagon are the local estimates by Fukugita et al. (1998) of the 
stellar mass density in spheroids, disks and irregulars, respectively.}
\end{figure*}

 \subsection{The evolution of the stellar mass density} \label{bozomath}
 Figure 9 shows the predicted and observed evolution of $\Omega_{*}$, namely 
 the stellar mass density (living stars plus remnants) divided by the critical density of the universe. 
 The stellar mass density is given by:
 \begin{equation}
 \rho_{*}(z) = \sum_{i} \rho_{B i}(z) \, (\frac{M}{L})_{B i}(z) 
 \end{equation} \\
where $ \rho_{B i}(z)$ is the B luminosity density and $(\frac{M}{L})_{B i}$ is the B mass-to-light ratio 
for the galaxy of the $i-$th morphological type.
In figure 9 we show two values of the stellar mass density measured in the local universe: the one by 
Cole et al. (2001), valid for the whole galaxy population, and
the one by P\'erez-Gonz\'alez et al. (2003), repesenting the contribution brought by local star forming
galaxies. The discrepancy between our results and the observations (and among the observations) 
at $z=0$ is likely to be ascribed to uncertainties in the normalization chosen for the local LF.
We notice that our prediction for the total stellar mass density is consistent with the 
values by Brinchmann \& Ellis (2000) and with the low redshift value by Dickinson et al. (2003). 
In our scenario, ellipticals build up the totality of their stellar mass at very early times, whereas 
disks of spirals and irregulars build up their stars progressively. Thus, the rise in the $3 \ge z \ge 0.5$ 
redshift range, observed by Dickinson et al. (2003), is mainly determined by spiral galaxies, namely by the 
same galaxies 
contributing to the drop in the cosmic SFR density between $z\sim1$ and $z=0$.\\
Finally, it is worth noting that our $z=0$ mass density values are in 
good agreement with the estimates by Fukugita et al. (1998) for spheroids, disks and irregulars.   

 \subsection{The missing metals crisis} \label{bozomath}
 The metals in the high redshift universe are observed in various amounts in different sites.
 One is represented by the DLA systems, which are considered to be the high-redshift counterparts 
 of the gas-rich 
 galaxies in the local universe and are usually associated with slowly-evolving galactic or sub-galactic 
 systems (Pettini et al. 1999, Prochaska et al. 1999, Calura et al. 2002).
 Other sites of significant metal production at high redshift are the Lyman-break galaxies (LBG, Steidel
 et al. 1996,
 Steidel et al. 1999) 
 whose high metal abundances, strong luminosities and kinematical features indicate their
 possible association with star-forming spheroids (Matteucci \& Pipino 2002, Pettini et al. 2002).
 A certain amount of the metals produced in proto-galaxies is ejected through SN-driven winds
 into the inter-galactic medium, whose presence is detected as a ``forest'' of absorption lines 
 bluewards of the Lyman-$\alpha$ emission lines of high redshift QSOs (for a review of the argument, see
 Pettini 2000,
 Bechtold 2001).
 The inter-galactic gas is highly ionised and is likely to account for most of the baryons at low and high
 redshift.
 In this medium the lowest metallicity in the universe can be observed (Songaila 2001), possibly due to
 enrichment 
 by pop III stars (Ostriker \& Gnedin 1996).\\
 Finally, an uncertain amount of metals is locked up in QSO and active galactic nuclei (Hamann 1997), 
 objects with metallicities even higher than the ones observed in LBGs (Matteucci \& Padovani 1993).\\
 We evaluate the metal ejection rate (MER) density according to Pagel (2001) by assuming the
 proportionality between the comoving density of metals produced by stars in the universe and the 
 SFR density to be a factor $1/42$ (Madau et al. 1996, Pettini 1999, Pagel 2001).
 This is in principle a rough approximation, since it is 
 equivalent to ignore the contribution of type Ia SNe and consider 
 only the elements produced explosively by massive stars.
 However, for our purposes this assumption is reasonable since 
 the main contributor to the total metallicity Z is the oxygen, 
 which is produced on very short timescales exclusively by massive stars. 
 A more refined calculation of the production rate in the universe of 
 various elements will be the object of a forthcoming paper (Calura \& Matteucci, in preparation). 

 If we integrate the predicted MER density across the epoch ranging from 13.3 and 11 Gyr, which
 corresponds, for the 
 $\Lambda$CDM cosmology, to the redshift range where the bulk of DLAs and LBG are found,
 we obtain:\\
 \begin{equation} 
 \int_{11Gyr}^{13.3 Gyr} \dot{\rho_{Z}}(t)dt = 5.2 \times 10^{6} M_{\odot} Mpc^{-3} \\
 \end{equation}
 An approximative estimate based on the observed values of $\rho_{*}$ yields $5 \times 10^{6} M_{\odot}
 Mpc^{-3}$ (Pagel 2001), 
 a value in excellent agreement with our appraisal, though slightly lower. 
 The observed contribution from DLAs and LBGs represents $\sim10\%$ of the estimate by Pagel and us. 
 We find such a high value for the total metal density at high redshift 
 since we assume that all ellipticals form at the same time.
 If we spread the formation of ellipticals over a wide redshift range, 
 we would probably find a lower value both for the SFR density peak and for 
 the metal density. For this reason, the values predicted for such peak and
 for the total amount of missing metals at high redshift should be regarded 
 as upper limits.
 However, this does not mean that the missing metal crisis could not 
 be even more serious than what the observations indicate if most 
 of the metals were produced in dust-obscured starbursts associated with early spheroid formation.
 We suggest that, irrespective of the amount of metals, the most plausible site of the missing metals
 could be the warm gas in galaxy groups 
 and proto-clusters, in which the metals could have been ejected through strong winds following the
 intense starbursts
 in ellipticals.
 In such an environment, the presence of diffuse metals could be difficult to detect, likely owing to
 virial temperatures lower
 than the ones observed in clusters (Renzini 1997, Fukugita et al. 1998).
 The main sources of strong winds could be the LBGs, where recent kinematic studies have indicated the
 frequent presence
 of strong large-scale outflows (Pettini et al. 2001, 2002).
 Another possible source is represented by the SCUBA (Blain et al. 1999, Trentham, Blain \& Goldader 1999)
 and hyperluminous infrared galaxies (Rowan-Robinson 2000), very luminous in the infrared band but faint
 in the optical,
 where the most intense observed starbursts are likely to occur and which are natural candidates for
 massive 
 proto-ellipticals (Ivison et al. 2000).

 \subsection{The cosmic SN rate} \label{bozomath}
 \begin{figure*}
 \centerline{\includegraphics[height=25pc,width=25pc]{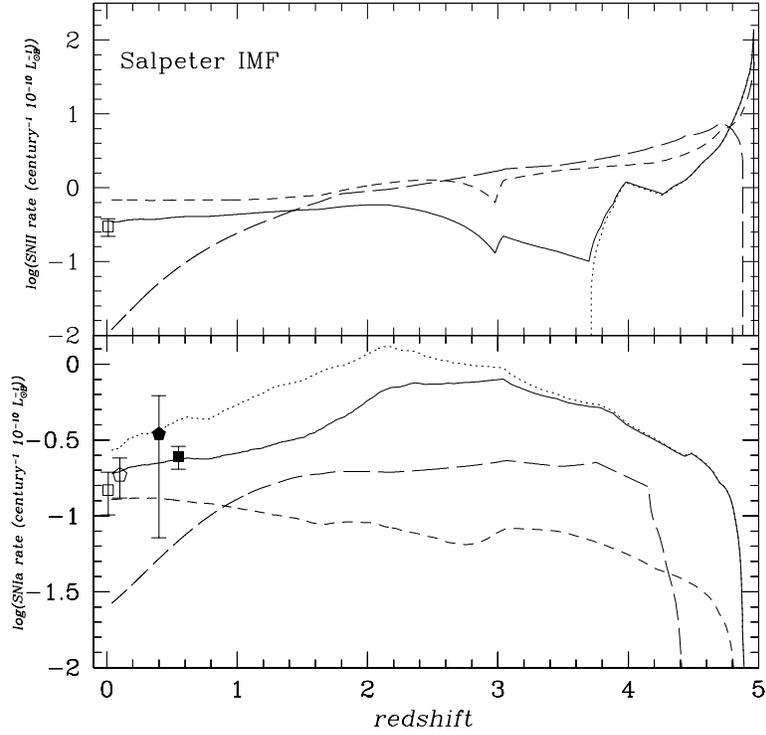}} 
 \caption[]{Predicted and observed type Ia (below) and II (above)
 SN rate expressed in SNu (SN century$^{-1}$
 $10^{-10} L_{\odot B}$) for a Salpeter IMF, a $\Lambda$CDM cosmology and galaxy formation at $z_{f}=5$.
 Dotted line: elliptical galaxies; short-dashed line: spiral galaxies; long-dashed line: irregular
 galaxies;
 solid line: total SN rate.
 Open squares: Cappellaro et al. 1999; open pentagon: Hardin et al. 2000;
 solid pentagon: Pain et al. 1996; solid square: Pain et al. 2002.
 }
 \end{figure*}

 \begin{figure*}
 \centerline{\includegraphics[height=25pc,width=25pc]{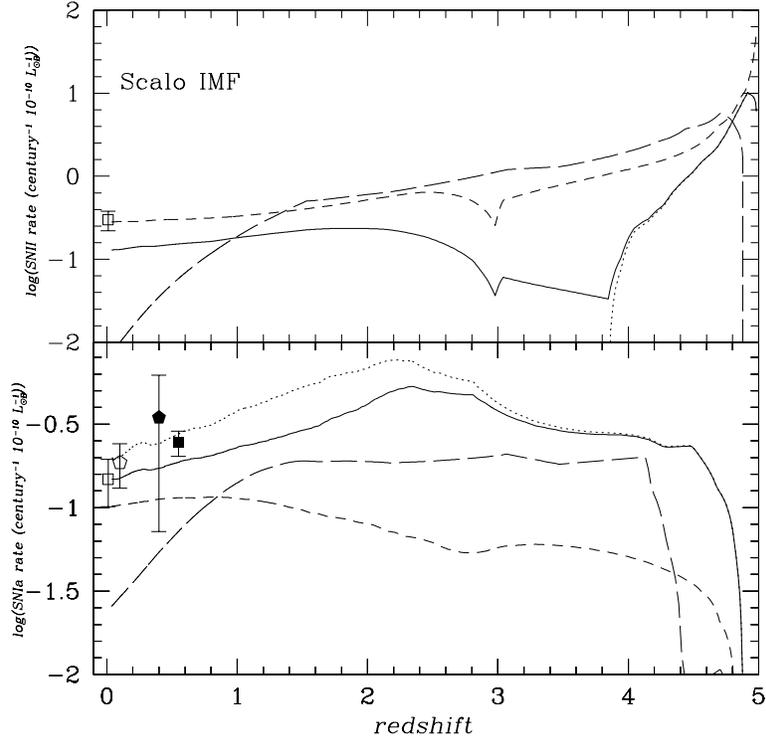}} 
 \caption[]{Predicted and observed type Ia (below) and II (above) SN rate 
 expressed in SNu (SN century$^{-1}$
 $10^{-10} L_{\odot B}$) for a Scalo IMF, a $\Lambda$CDM cosmology and galaxy formation at $z_{f}=5$.
 Curves and symbols are as indicated in figure 10.
 } 

 \end{figure*}

 Another key-issue in the study of galaxy evolution is the determination of the cosmic SN rate,
 which can provide useful constraints on 
 the cosmological parameters (Marri, Ferrara \& Pozzetti 2000), 
 on the possible SN progenitor models and on the evolution of the 
 cosmic star formation history.
 The observed cosmic SN rate is expressed in SNu, i. e. in SN per unit time per unit blue luminosity:\\
 \begin{equation}
 1 SNu=1SN/10^{10}L_{\odot B}/century
 \end{equation}\\
 The observed SN rate $R$ is expressed as (Hardin et al. 2000):
 \begin{equation}
 R=N/S
 \end{equation}\\
 where N is the number of SN detected and S is a sum over the observed galaxies weighted
 by their blue luminosities $L_{i}$:\\
 \begin{equation}
 S=\sum_{i} L_{i}\int^{\infty}_{-\infty}\epsilon_{i}(t,z_{i}) dt
 \end{equation}\\
 where $\epsilon$ is the efficiency to detect in the $i-$th galaxy a SN whose maximum 
 occurs at time t in the SN rest frame.
 For similarity, we calculate the cosmic SN rate at redshift z as:
 \begin{equation}
 R_{c}(z)=\frac{\sum_{i} r_{i}(z)}{\sum_{i}L_{B i}(z)}
 \end{equation}\\
 where $r_{i}(z)$ represents the number of SNe per 100 yr exploding in the $i-$th galaxy type at redshift
 z,
 whereas $L_{B i}$ is the predicted blue luminosity of the $i-$th galaxy type at redshift z.
 Figures 10 and 11 show the results of our computation of the cosmic type Ia and II
 SN rates, calculated in SNu (1 SNu= 1 SN per century per $10^{10} L_{\odot B}$), having assumed
 a $\Lambda$CDM world model and galaxy formation occurring at $z_{f}=5$.
 In figure 10 we adopt a universal Salpeter IMF, whereas in figure 11 a Scalo IMF.
 We compare our predictions for galaxies of different morphologies with data observed 
 by several authors up to $z\sim0.55$ (see the captions to figures 10 and 11 for details).
 According to our calculations, in the local universe the bulk of SNe Ia explode in elliptical and spiral
 galaxies. 
 At increasing redshift the contribution of ellipticals becomes more and more significant.
 At $z\sim0$, practically all type II SNe explode in spiral galaxies, with a small contribution brought by
 irregular galaxies. 
 In ellipticals the type II SNe explode only at high redshift, since these galaxies are 
 assumed to evolve passively since the interruption of the star formation at the onset of the galactic
 wind, 
 which occurs at early times.
 We note that our fit to the observed points is excellent in the case of the Salpeter IMF both for type Ia
 and II SNe.
 In the case of the Scalo IMF we underestimate the number of SN II in the local universe and we slightly
 underpredict the type 
 Ia SN rate observed at high redshift ($z\sim0.55$, Pain et al. 2002).
 We stress that, unlike all our previous predictions, the calculation of the SN rate is independent on the
 luminosity function 
 and is directly comparable to the observations without requiring a normalization at $z=0$.
 For this reason, the calculation of the SN rate can be considered a very reliable test of consistency for
 our 
 chemo-spectrophotometric model.\\
 If the cosmic SN rate can be considered as a reliable indicator of how star formation has evolved in the
 universe since the 
 growth of galactic structures, as discussed in Madau, Della Valle \& Panagia (1998) and Sadat et al.
 (1998), our result is a 
 further evidence that the bulk of the galaxy population has evolved in luminosity and not in number since
 early times, 
 i.e. that the bulk of the galaxies as we see them today were already in place at early epochs.
 Observations of the cosmic SN rate at very high redshifts, 
 hopefully achievable after the launch of the \emph{Next Generation Space Telescope} (NGST),
 can provide fundamental clues on the galaxy formation epoch.

 \section{Conclusions}
 In this paper we have used detailed chemo-spectrophotometric models for galaxies of different
 morphological types,
 namely ellipticals, spirals and irregulars, in order to study the evolution of the galaxy luminosity
 density in various bands,
 the cosmic SFR density and the cosmic SN rate. 
 We have studied the evolution of the 
 baryonic matter in the universe in the form of stars and metals, starting from the B-band luminosity
 function 
 observed in the local universe by Marzke et al. (1998), having assumed 
 that the whole galaxy population started forming stars at high redshift.
 We have considered two different forms for the universal IMF, namely the Salpeter (1955) and the Scalo
 (1986), and 
 we have computed our models for different cosmologies, i. e. the Einstein-De Sitter and the $\Lambda$CDM.
 We have 
 compared our predictions with an
 updated large set of high-redshift observations and we found an overall good agreement 
 between our predictions and all the observational evidence considered in the present work.
 In particular, our main results are:\\
 1) We reproduce the evolution of the galaxy luminosity density in the B, I and K bands 
 considering that all galaxy types evolve in luminosity and not in number since the epoch of their
 formation. Generally, we overestimate the luminosity density observed in the UV, in particular 
 at high redshift. 
 In our scheme, elliptical galaxies suffer high-redshift starbursts after which they 
 evolve as passive systems, i. e. without star formation, whereas spiral disks 
 and irregular galaxies form stars continuously down to the present epoch.
 At high ($>2$) redshift, ellipticals dominate the total emissivity in any optical band.
 We predict a high redshift peak generated by the intense starburst in ellipticals which
 is not visible in the available observations at short wavelengths not corrected by extinction.
 The cause could be a heavy attenuation by interstellar dust which could hide the most intense 
 star-forming sites, or the bulk of the rest-frame UV light emitted by single objects. 
 On the other hand, extinction-corrected high redshift measures of the UV flux density (Massarotti et al.
 2001) 
 are fully compatible with our predictions.
 High-redshift observations in bands less sensitive to dust attenuation, currently absent, 
 would be fundamental in probing the existence of such a peak.
 At $z<2$, in the U and B bands the disks of spiral galaxies emit the bulk of the light and 
 are the main contributors to the decline in the galaxy luminosity density 
 observed between $z\sim1$ and $z=0$, whereas in the the I and K bands the light emitted 
 by stars in ellipticals is comparable to that emitted by the spirals. Irregular galaxies bring a 
 negligible contribution to the total luminosity density at any epoch and in any band.
 Both the Salpeter and Scalo IMFs provide very satisfactory results, 
 especially assuming a $\Lambda$CDM cosmological model.\\
 2) 
We predict that spiral disks are the most active star forming sites in the universe throughout the redshift 
interval $0<z<1$ which, for a standard $\Lambda$CDM cosmology and $h=0.65$, correspond to the $57\%$ of the cosmic time. 
Furthermore, if the SFRs recorded in SCUBA and hyperluminous infrared galaxies can be considered typical
for starbursting protoellipticals, we predict a very high peak (up to $\sim 1000$ times the value observed in 
the local universe) in the cosmic SFR density occurring at the epoch when the bulk of the spheroids 
formed.\\
 3) According to our calculations, the missing metal crisis could be even more serious than what has been
 proposed 
 by previous authors (Pettini 1999, Pagel 2001) if the bulk of the metals were produced in dust-obscured
 starbursts 
 associated with an early spheroid formation.
 We suggest that, regardless of the amount of metals, the most plausible site for the missing metals could
 be the 
 warm gas in galaxy groups 
 and proto-clusters, in which the metals could have been ejected through strong winds following intense
 starbursts. 
 In such an environment, the presence of the metals would be difficult to detect owing to the low virial
 temperatures.
 In principle, it could be possible to detect these metals by identifing 
 the most recently assembled clusters of galaxies, provided that they 
 have formed out of groups where the missing metals lie.
 The main sources of such strong winds could be both Lyman-break galaxies (Pettini et al. 2002), SCUBA
 (Trentham et al. 1999)
 and hyperluminous infrared galaxies (Rowan-Robinson 2000), which are the best candidates for the
 high-redshift 
 counterparts of the nearby massive elliptical galaxies.\\
 4) The study of the cosmic SN rate is a fundamental test of consistency for our models since it is
 completely independent 
 of the normalization of the galaxy population at $z=0$, which was required by all our previous
 calculations.
 We reproduce very well the observed cosmic SN rate assuming a universal Salpeter-like IMF.
 The assumption of a Scalo IMF causes an underestimate of the local type II SN rate and 
 of the Ia SN rate observed at high redshift.
 We predict that most of the SN Ia explode in ellipticals at all ages, whereas all the SN II explode in
 spiral disks and irregulars
 after the halt of star formation in elliptical galaxies occurred at high redshift.
 This is in fair agreement with all our previous results, which indicate that spiral galaxies 
 are the most active star forming objects since at least $z\sim1$. 
 Also the observed cosmic SN rate can be explained by assuming a scenario of coeval galaxies whose number 
 is conserved and which evolve purely in luminosity. 
 The determination of the cosmic SN rate at redshifts $>0.55$ (i.e. the highest explored by current SN
 surveys,  
 Pain et al. 2002), hopefully achievable after the launch of the \emph{Next Generation Space Telescope}, 
 can provide further constraints on galaxy evolution models.\\
 All the results obtained in this paper clearly indicate that the bulk of galaxies was already 
 in place at early epochs. According to several recent observations, the presence of high mass galaxies at
 high redshift 
 is not rare (Saracco et al. 2002, Rusin et al. 2002, Pozzetti et al. 2003). 
 Furthermore, in the last few years increasing evidence has been 
 found that most (up to 70\%) of the extremely red objects discovered in deep NIR and optical surveys
 could be 
 high-redshift ellipticals (Daddi et al. 2000 and references therein).
 Certainly, all these evidences seem difficult to reconcile with galaxy 
 formation scenarios where the bulk of the stellar mass is formed at $z<1$.

 \acknowledgments

 We are grateful to Raul Jimenez for having provided us with the 
 spectrophotometric code and to Andrea Biviano for providing the code 
 to calculate the fit to the Schechter function.
 We thank John Danziger 
 and C\'eline P\'eroux for a very careful reading of the manuscript
 and for very useful comments, and A. Pipino, C. Chiappini, L. Silva and 
 P. Monaco for many interesting discussions.
 Very special thanks go to Lucia Pozzetti for many valuable 
 suggestions on an early version of the paper.
 Finally, we wish to thank an anonymous referee for his enlighting comments.

 \end{document}